\documentclass[12pt]{article}
\usepackage{epsf}
\usepackage{graphicx}
\usepackage{eepic}
\usepackage{latexsym}
\usepackage{amsmath}
\newcommand{\be}{\begin{equation}}
\newcommand{\ee}{\end{equation}}
\newcommand{\ba}{\begin{array}}
\newcommand{\ea}{\end{array}}
\newcommand{\bqa}{\begin{eqnarray}}
\newcommand{\eqa}{\end{eqnarray}}
\begin{document}
\begin{center}
{\Large\bf  Unitarity constraints on chiral perturbative amplitudes
}
\\[10mm]
{\sc  H.~Q.~Zheng\footnote{Talk given at: International Workshop on
Effective Field Theories: from the pion to the upsilon,
        February 2-6 2009,  Valencia, Spain}}
\\[2mm]
{\it  Department of Physics, Peking University, Beijing 100871,
P.~R.~China }
\begin{abstract}
Low lying scalar resonances emerge as a necessary part to adjust
chiral perturbation theory to experimental data once unitarity
constraint is taken into consideration. I review recent progress
made in this direction in a model independent approach. Also I
briefly review studies on the odd physical properties of these low
lying scalar resonances, including in the
$\gamma\gamma\to\pi^+\pi^-, \pi^0\pi^0$ processes.
\end{abstract}
\end{center}


 Low lying scalar resonances emerge as a necessary part to
adjust chiral perturbation theory to experimental data when the
constraint of unitarity is  taken into consideration. This is most
clearly seen if one writes down a dispersion relation for $\sin
(2\delta_\pi)$ where $\delta_\pi$ is the $\pi\pi$ scattering phase
shift in the scalar--iso-scalar channel. The data exhibits a convex
curvature below 1GeV whereas chiral estimation to the nearby cut
contribution is negative and concave -- the huge gap between the two
can only be made up by including a pole contribution, according to
the standard $S$--matrix theory principal.~\cite{XZ00} Unitarization
of the chiral perturbative amplitude  also predicts the existence of
a light and broad pole structure in the IJ=00 channel $\pi\pi$
scattering,
 nevertheless it was not very clear to what
extent one should trust the output of unitarized chiral perturbative
amplitude.

In section~\ref{PKU} we briefly introduce a novel dispersion
representation for partial wave amplitudes developed in recent few
years,~\cite{zhou04,zhou05} and physical results read out from it,
including a better understanding on the Pad\'e unitarization
approximation. In section~\ref{matching} we discuss how one can get
a better understanding on chiral perturbation theory and  resonance
chiral perturbation theory (R$\chi$PT) parameters based on the use
of dispersion techniques. In section~\ref{f0600} we investigate
studies on the dynamical properties of the low lying scalar
resonances. Finally in section~\ref{gammapipi} we introduce a recent
work on the $\gamma\gamma\to\pi\pi$ process. Based on which we find
that the $\sigma\to\gamma\gamma$ coupling is significantly smaller
than that of a naive $\bar qq$ assignment.

\section{The PKU representation -- a unitarized dispersion
representation for elastic scattering amplitudes }\label{PKU}

The $S$-matrix element of partial wave elastic scattering amplitude
satisfies the following dispersive
representation:~\cite{zhou04,zhou05}
\begin{equation}\label{Separable}
 S^{phy.}=\prod_iS^{R_i}\cdot S^{cut}\ ,
\end{equation}
where $S^{R_i}$ denotes the $i$-th $second$ sheet pole contribution
and $S^{cut}$ denotes the contribution from cuts except the elastic
one. The information from higher sheet poles is hidden in the right
hand integral which consists of one part of the total background
contribution. We have,
\begin{eqnarray}\label{fs}
 S^{cut}&=&e^{2i\rho f(s)}\ ,\nonumber\\
 f(s)&=&\frac{s}{\pi}\int_{L}\frac{{\rm
 Im}_Lf(s')}{s'(s'-s)}+\frac{s}{\pi}\int_{R}\frac{{\rm
 Im}_{R}f(s')}{s'(s'-s)}\ ,
\end{eqnarray}
where the `left hand' cut $L=(-\infty, 0]$ for equal mass
scatterings and may contain a rather complicated structure for
unequal mass scatterings. The right hand cut $R$ starts from first
\textit{inelastic} threshold to positive infinity. It can be
demonstrated that the dispersive representation for $f$ is free from
the subtraction constant.~\cite{zhou04} The PKU representation,
Eq.~(\ref{Separable}), is sensitive to $S$ matrix poles not too far
away from physical threshold, hence providing a useful tool to
explore the light and broad resonance $\sigma$ and $\kappa$. In the
data fit it is found that crossing symmetry  plays an important role
in fixing the $\sigma$ pole location. Taking this fact into
account~\cite{zhou05} it gives the $\sigma$ pole location at $
M_\sigma=470\pm 50\mbox{MeV}\ ,\,\,\, \Gamma_\sigma=570\pm
50\mbox{MeV}\ , $ in good agreement with the determination using
more sophisticated Roy equation analysis.~\cite{caprini,pelaez2} The
application of Eq.~(\ref{Separable}) to LASS data~\cite{LASS} also
unambiguously establish the existence of the $\kappa$ meson with the
pole location:~\cite{zhou04} $ M_\kappa=694\pm 53\mbox{MeV}\ ,\,\,\,
\Gamma_\kappa=606\pm 59\mbox{MeV}\ , $ which are also in agreement
with the  later determination on $\kappa$ pole parameters using
Roy--Steiner equations.~\cite{descotes}

The dispersion representation Eq.~(\ref{PKU}) safely embeds chiral
perturbative amplitudes into a unitarized scheme. This property is
not always trivial in the practice of unitarization. For example,
contrary to the input chiral perturbative amplitudes, Pad\'e
approximants lead to completely different singularity structure in
the vicinity of $s=0$ -- a region where the former ought to be
trustworthy. The reliability of $\chi$PT predictions in the small
$|s|$ region can be vividly seen in the I=2 $s$ wave amplitude. One
may use $\chi$PT result to estimate the contribution of the left
hand cut to the scattering length $a^2_0$. The estimate is rough but
gives a value qualitatively much larger in magnitude than $a^2_0$
extracted from experiment. This difficulty is resolved, recalling
that $\chi$PT also predicts a virtual pole near $s=0$, which
contribution cancels a large amount of the cut contribution and
leads to the correct prediction of $a^2_0$. This example further
illustrate that the singularity structure as predict by $\chi$PT in
the vicinity of $s=0$ is indeed reliable and self-consistent.

 The PKU representation, Eq.~(\ref{Separable}) affords
another interesting opportunity to study the relation between
resonance parameters and low energy constants of the chiral
lagrangian. On the $l.h.s.$ of Eq.~(\ref{Separable}) one may replace
$S^{phy.}$ by $\chi$PT result at low energies, on the $r.h.s.$ one
does not know how many resonances are there, nevertheless one may
formally make a threshold expansion to match the $l.h.s.$. In this
situation the cut integrals in Eq.~(\ref{fs}) are however difficult
to calculate. So we firstly neglect completely the cut integrals and
assuming there is only one pole on the $r.h.s.$ of
Eq.~(\ref{Separable}). Under this approximation we can get a
prediction on the resonance parameters expressed in terms of the low
energy constants. It is shown in Ref.~\cite{sun} that the pole
location is exactly the same as the prediction of [1,1] Pad\'e
amplitude in the large $N_c$ and chiral limit. In my knowledge there
had never been serious examination on what does the predictions of
Pad\'e approximation mean in the literature. Hence Ref.~\cite{sun}
provides a first understanding on this question: in the large $N_c$
and chiral limit the pole location as predicted by [1,1] Pad\'e
approximant is equivalent to the approximation that: 1) neglecting
crossed channel cut completely, 2) assuming single pole dominance in
the $s$ channel. However, in Ref.~\cite{juanjo07} it is shown that
crossed channel cut contributions are \textit{not} negligible. A
direct consequence of neglecting the cut contribution is the
violation of crossing symmetry as will be discussed in the next
section.
\section{Matching between two expansions}\label{matching}

Threshold expansion on both sides of Eq.~(\ref{Separable}) (the
$l.h.s.$ replaced by $S^{\chi PT}$) could provide useful relations
between resonance parameters and low energy constants, if one can
reliably estimate cut integrals in some way. Fortunately, this can
indeed be done in the large $N_c$ limit. In such a limit the
Eq.~(\ref{Separable}) leads to the same result as the partial wave
dispersion relation. Hence Eq.~(\ref{Separable}) can actually be
understood as a simple combination of single channel unitarity and
 partial wave dispersion relation.~\cite{juanjo07}

The matching results in a set of relations at different chiral
order.\\

 \vspace{.5cm}
 \noindent {\bf At  $O(p^2)$:}
\begin{equation}
\label{sumksrf} {\frac{1}{16 \pi f^2}=\frac{9\Gamma^{(0)}_V}{M_{\rm
V}^{(0)\, 3}}}+\frac{2\Gamma^{(0)}_S}{3M_{\rm S}^{(0)\,
3}},\nonumber
\end{equation}
It is remarkable to notice that three different channels produce the
same results. The conclusion is that \textit{partial wave amplitudes
remember crossing symmetry.} It is interesting to compare
Eq.~(\ref{sumksrf}) with the old version of the so called KSRF
relation:
\begin{equation}\label{KSRF}
\frac{1}{16 \pi f^2}=\frac{6\Gamma^{(0)}_V}{M_{\rm V}^{(0)\, 3}}\ .
\end{equation}
In the IJ=11 channel, one may obtain Eq.~(\ref{KSRF}) if  neglecting
crossed channel vector and scalar exchanges, see
table~\ref{tab.KSRF} for illustration.~\cite{juanjo07}
\begin{table}[!t]
\begin{center}
\begin{tabular}{|c|c|c|c|c|}
  \hline
\rule[-0.7em]{0em}{1.9em}
   & $T(0)$ & $t_0^{\rm tR}$ & $t_0^{\rm sR}$ & $t_0^{\chi PT}=m_\pi a_{J}^I$ \\
  \hline
\rule[-0.7em]{0em}{1.9em}
  $IJ=11$ & $-\frac{m_\pi^2}{24\pi f^2}$ & $\frac{4\Gamma_S}{9M_S^3}+\frac{2\Gamma_V}{M_V^3}$
        & $\frac{4\Gamma_V}{M_V^3}$ & 0 \\
  \hline
\rule[-0.7em]{0em}{1.9em}
  $IJ=00$ & $-\frac{m_\pi^2}{32\pi f^2}$ & $-\frac{4\Gamma_S}{3M_S^3}+\frac{36\Gamma_V}{M_V^3}$
        & $\frac{4\Gamma_S}{M_S^3}$ & $\frac{7m_\pi^2}{32\pi f^2}$\\
  \hline
\rule[-0.7em]{0em}{1.9em}
  $IJ=20$ & $\frac{m_\pi^2}{16\pi f^2}$ & $-\frac{4\Gamma_S}{3M_S^3}-\frac{18\Gamma_V}{M_V^3}$
        & 0 & $-\frac{m_\pi^2}{16\pi f^2}$ \\
  \hline
\end{tabular}
\caption{{\small Summary of the different contributions  $T(0)$,
cross channel resonance exchange contribution $t_0^{\rm tR}$, and
$s$-channel resonance contribution $t_0^{\rm SR}$ to the scattering
lengths at leading order in the $m_\pi^2$ expansion. The generalized
KSRF-relation derives from the matching of the sum of the first
three columns to the $\chi$PT prediction, $t_0^{\chi PT}$. In the
last line,  $T(0)$ contains the sum of $-|T(0)|$ and the $IJ=20$
virtual pole contribution. }} \label{tab.KSRF}
\end{center}
\end{table}
\\ \vspace{.5cm}

\noindent {\bf At $O(p^4)$:}
\begin{eqnarray}
\label{sumL2} L_2\, = \, 12\pi f^4\frac{\Gamma_{\rm V}^{(0)}}{M_{\rm
V}^{(0)\, 5}} \ ,  \,\,\,\,\,\,  L_3 = 4 \pi f^4
\left(\frac{2\Gamma_{\rm S}^{(0)}}{3M_{\rm S}^{(0)\,
5}}-\frac{9\Gamma_{\rm V}^{(0)}}{M_{\rm V}^{(0)\, 5}}  \right)\, .
\end{eqnarray}
It is remarkable to notice that the Eq.~(\ref{sumL2}) rewrites the
old results of Ref.~\cite{ecker} without even knowing how to write
down an effective resonance chiral lagrangian!
\\

\vspace{.5cm} \noindent {\bf At $O(m_\pi^2p^2)$:}\\

 Matching at this order led to a novel relation
any lagrangian model has to obey, which is a consequence of high
energy constraint combined with chiral symmetry:
\begin{eqnarray} \label{sumalpha} 0&=& \frac{2}{3}\,\,
\frac{\Gamma_{\rm S}^{(0)}}{M_{\rm S}^{(0)\, 5}}\,
\left[\alpha_S+6\right] \, +\, \frac{ 9\, \Gamma_{\rm
V}^{(0)}}{M_{\rm V}^{(0)\, 5}} \, \left[\alpha_V+6\right]\, .
\end{eqnarray}
The physical widths and masses, $\Gamma_{\rm R}$ and $M_{\rm R}$,
carry an implicit dependence on $m_\pi^2$, which can be expressed in
the form
\begin{equation}
\frac{\Gamma_{\rm R}}{M_{\rm R}^3}\, =\, \frac{\Gamma_{\rm
R}^{(0)}}{M_{\rm R}^{(0)\, 3}}\,\left[\, 1 \, +\, \alpha_R\,
\frac{m_\pi^2}{M_{\rm R}^{(0)\, 2}}\,\,\, +\, \, \,
\mathcal{O}\left(m_\pi^4\right)\, \right]\, .
\end{equation}
The matching project has been  further extended to $O(p^6)$ and
interesting results are obtained,~\cite{GSZ2} I refer to the talk
given by Sanz--Cillero in this conference for details. Based on
these new formulas, Guo and Sanz--Cillero made a systematic
re-estimation to the coupling constants in $O(p^6)$ chiral
lagrangian in a model independent way.~\cite{GS09}

\section{On the nature of the f0(600) pole}\label{f0600}

It has long been argued that the $f_0(600)$ pole is the $\sigma$
meson of linear $\sigma$ (-like) model.~\cite{sigma} Nevertheless
due to its strong interaction nature, it is very difficult to solve
this problem at fundamental level. On the other side, it is argued
using the inverse amplitude method or chiral unitarization approach,
that the $f_0(600)$ pole is a `dynamically generated'
resonance.~\cite{pelaez3} Yet the wording `dynamically generated'
itself needs clarification,~\cite{giacosa} it can be understood from
the discussion in section~\ref{PKU} and Refs.~\cite{XZxx,sun} that
in the approach of unitarization of chiral perturbative amplitudes
the $f_0(600)$ pole does fall back to the real axis in the large
$N_c$ limit. Hence one has to put it explicitly in the lagrangian in
the very beginning, therefore being `fundamental'. The odd pole
trajectory of $f_0(600)$ with respect to the variation of $N_c$ was
used to argue its dynamical nature. Nevertheless it is shown, using
a solvable $O(N)$ linear $\sigma$ model, that the `fundamental'
$\sigma$ pole trajectory looks indeed being odd.~\cite{Xiaoly} One
may expect that the study of other light scalar resonances like
$\kappa(700)$, $f_0(980)$ and $a_0(980)$  could shed some light on
the understanding of $f_0(600)$. However, the inclusion of these
resonances does not seem to be very helpful up to now, if not merely
making the situation more complicated.

\section{
The $\gamma\gamma\to \pi\pi$ process in a partially couple channel
approach}\label{gammapipi}

\begin{figure}[h]
\centering
\includegraphics[height=6.5cm,width=14cm]{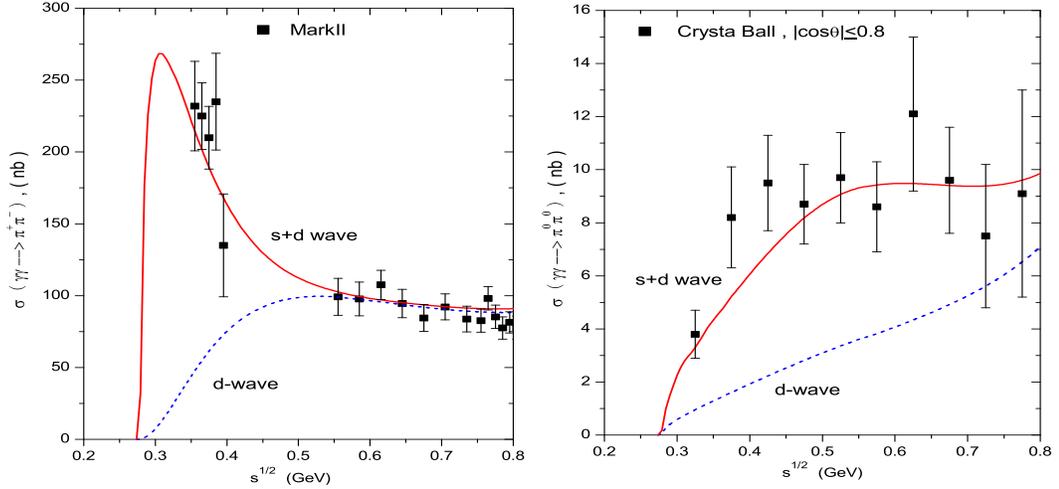}
\caption{\label{fig}A fit up to 0.8GeV using single channel $s$-wave
$T$ matrices of Ref.~\cite{zhou05}, with only one fit parameter. The
$\pi^+\pi^-$ and $\pi^0\pi^0$ data are from Refs.~\cite{Boyer1990}
and \cite{Marsiske1990}, respectively. Dashed curve represents
$d$-wave background, solid curve represents the total contributions,
including the I=0 $s$--wave to be fitted.
}
\end{figure}

There remains  the hope in understanding better the property of
$f_0(600)$ through the study on the $\gamma\gamma\to
\pi^+\pi^-,\pi^0\pi^0$ process, as emphasized by
Pennington,~\cite{pennington} since the di-photon coupling of a
resonance may be used as probe to investigate hadron internal
structure at quark level. Again unitarity plays a crucial role in
such investigations. However the $\sigma$ pole locates quite far
away from the physical region, the di-photon coupling extracted as
such is found not very stable. This problem is reinvestigated
recently,~\cite{mao} where the fit to data at first step is up to
1.4GeV, aiming at fixing the $d$-wave background. Then a refined
analysis is made by fitting data up to 0.8GeV, using the $\pi\pi$
scattering $T$ matrix obtained in Ref.~\cite{zhou05}. The fit
quality can seen in fig.~\ref{fig} borrowed from Ref.~\cite{mao}.
In this way the two photon decay width is obtained to be
$\Gamma(\sigma\to\gamma\gamma)\simeq 2.1$KeV, a result significantly
smaller than that expected from a naive $\bar qq$ model calculation.
This further stresses the unconventional nature of the $f_0(600)$
nature except the  large width it has. The value of $\sigma \pi\pi$
coupling is also given in Ref.~\cite{mao},
$g^2_{\sigma\pi\pi}=(-0.20-0.13i)\mathrm{GeV}^2$, which is
compatible with the value given in Ref.~\cite{menn}:
$g^2_{\sigma\pi\pi}=-0.25-0.06i\mathrm{GeV}^2$ which was used by the
authors of Ref.~\cite{kaminsky} to argue in favor of the gluonium
nature of $f_0(600)$. It is interesting to notice that, for a narrow
resonance $\mathrm{Re}[g^2_{\sigma\pi\pi}]$  should be positive,
otherwise it would be a ghost rather than a particle and violates
probability conservation. Nevertheless for a broad resonance this
constraint does not need to hold anymore. The negative
$\mathrm{Re}[g^2_{\sigma\pi\pi}]$ indicates another peculiarity of
$f_0(600)$.

To conclude, the correct use of unitarity, when combined with chiral
symmetry, plays a powerful  role in studying resonance physics,
especially the property of the light and broad $f_0(600)$. However,
there still remains many interesting and mysterious characters of
$f_0(600)$ waiting to be resolved in future.

\textit{\bf Acknowledgement:} The author would like to thank the
organizers of EFT09, especially Jorge Portoles for providing the
charming atomsphere for the conference and kind hospitality towards
him. This work is supported in part by National Nature Science
Foundation of China under Contract
Nos. 10875001 and 
10721063.


\begin{thebibliography}{99}
\bibitem{XZ00}Z.~G.~Xiao and H.~Q.~Zheng, Nucl. Phys. {\bf A695}
(2001)273; J.~Y.~He, Z.~G.~Xiao, H.~Q.~Zheng, Phys. Lett. {\bf
B536}(2002)59, Erratum-ibid. {\bf B549}(2002)362.


\bibitem{zhou04}Z.~Y.~Zhou, H.~Q.~Zheng, Nucl. Phys.{\bf
A775}(2006)212; H.~Q.~Zheng et al., Nucl. Phys. {\bf A733}(2004)235.

\bibitem{zhou05}Z.~Y.~Zhou et al., JHEP 0502(2005)043.


\bibitem{caprini}I.~Caprini, G.~Colangelo,
H.~Leutwyler, Phys. Rev. Lett. {\bf 96}(2006)132001.

\bibitem{pelaez2}R.~Garcia-Martin, R.~Kaminski, J.~R.~Pelaez,  Int.~J.~Mod.~Phys.~{\bf
A24}(2009)590 and references therein.

\bibitem{LASS}D.~Aston et al. (LASS Collaboration), Nucl. Phys. {\bf B296}(1988)493.

\bibitem{descotes}S.~Descotes-Genon, B.~Moussallam,
Eur. Phys. J. {\bf C48}(2006)553.


\bibitem{sun}Z.~X. Sun, L.~Y.~Xiao, Z.~G.~Xiao, H.~Q.~Zheng,
 Mod. Phys. Lett. {\bf A22}(2007)711.

 \bibitem{juanjo07}Z.~H.~Guo, J.~J.~Sanz--Cillero, H.~Q.~Zheng,
JHEP 0706(2007)030.

\bibitem{ecker}
    G. Ecker {\it et al.},
    {Phys. Lett.}  {\bf B223}(1989)425.

\bibitem{GSZ2}Z.~H.~Guo, J.J.~Sanz-Cillero, H.~Q.~Zheng, Phys. Lett. {\bf B661}(2008)342.

\bibitem{GS09}
Z.~H.~Guo, J.~J.~Sanz-Cillero,  e-Print: arXiv:0903.0782 [hep-ph].

\bibitem{sigma}N.~A.~Tornqvist, Z.~Phys.~{\bf C68}(1995)647; N.~N.~Achasov and
G.~N.~Shestakov, PRD 49(1994)5779; R.~Kaminski, et al., PRD
50(1994)3154; M.~Ishida et al.,  Prog. Theor. Phys. {\bf
99}(1998)1031; D.~Black et al., PRD 64 (2001) 014031.
 M.~D.~Scadron et al., Nucl. Phys. {\bf A724}(2003)391; M.~X.~Su, L.~Y.~Xiao, H.~Q.~Zheng, Nucl. Phys. {\bf
 A}792(2007)288.

\bibitem{pelaez3}
J.~R.~Pelaez, Phys.~Rev.~Lett.~{\bf 92}(2004)102001;
 Mod.~Phys.~Lett.~{\bf A19}(2004)2879;
 J.~R.~Pelaez, G.~Rios, Phys.~Rev.~Lett.{\bf 97}(2006)242002.
J.~A.~Oller et al., Phys. Rev. {\bf D59}(1999)074001.

\bibitem{giacosa}
F.~Giacosa,  e-Print: arXiv:0903.4481 [hep-ph].

\bibitem{XZxx}  Z.~G.~Xiao, H.~Q.~Zheng, Mod. Phys. Lett. {\bf A22}(2007)55.

\bibitem{Xiaoly} L.~Y.~Xiao, Z.~H.~Guo, H.~Q.~Zheng,  Int. J. Mod. Phys.
 {\bf A22}(2007)4603.

 \bibitem{pennington}
 M.~R.~Pennington, invited talk at YKIS Seminar on \textit{New Frontiers in QCD: Exotic Hadrons and Hadronic Matter},
 Kyoto, Japan, 20 Nov - 8 Dec 2006.
 Prog. Theor. Phys. Suppl. {\bf 168}(2007)143.

 \bibitem{mao}Y.~Mao et al.,
e-Print: arXiv:0904.1445 [hep-ph].

\bibitem{Boyer1990} J.~Boyer {et al}. (Mark-II Collaboration), Phys.~Rev. {\bf D42}(1990)1350.

\bibitem{Marsiske1990} H.~Marsiske {et al}.(Crystal Ball Collaboration), Phys. Rev. {\bf D41}(1990)3324.

\bibitem{menn}G.~Mennessier, S.~Narison, W.~Ochs, Phys. Lett. {\bf B665}(2008)205.
\bibitem{kaminsky}R.~Kaminski, G.~Mennessier, S.~Narison,
e-Print: arXiv:0904.2555 [hep-ph].
\end{thebibliography}
\end{document}